\documentclass[12pt,preprint]{aastex}

\slugcomment{Not to appear in Nonlearned J., 45.}

\received{2008 June 5; accepted 2008 September 18}
\begin{document}

\title{VARIATION OF THE LIGHT AND PERIOD OF THE MAGNETIC CATACLYSMIC VARIABLE AM Her\thanks{Based on observations gathered with the 1.5m Russian--Turkish telescope (RTT150) and ROTSE IIId at the T\"UB\.ITAK National Observatory.}}

%\author{B. Kalomeni\altaffilmark{2} and K. Yakut\altaffilmark{3,4}}
\author{Belinda Kalomeni \\
Department of Physics, \.Izmir Institute of Technology, 35430  \.Izmir, Turkey\\
and \\
Kadri Yakut\\
Institute of Astronomy, University of Cambridge, Madingley Road, Cambridge CB3 0HA, UK\\
Department of Astronomy \& Space Sciences, University of Ege, 35100  \.Izmir,  Turkey\\
}

\begin{abstract}
Ground-based long-term optic variability of AM Her,
covering the period between 2003-2008, has been conducted to study the features
seen in both low and high states of the system. Low-state analysis
shows the presence of short-term, low-amplitude light variations of
about 0.02-0.03 mag with a mean power time between 16 s and 226 s. Brightness variations on the order of 0.7--2 mag, which could be due to the stellar activity of the component in the system, are also detected.
A total of 30 years times of minimum light given in the literature are combined with nine times of minima obtained in this study. We represented the (observed--calculated) diagram by a parabolic curve and also by two broken lines. Under the assumption of a parabolic variation,
we estimate an increase in period, $\frac{dP}{dt}=7.5(1.2)\times10^{-9}$~days~yr$^{-1}$,
with a mass transfer rate of $\dot M = 8(2)\times10^{-9}$$M_{\odot} yr^-1$, in agreement with the
previous findings by a different method.
\end{abstract}


\section{Introduction}

AM Her systems (polars) are semidetached binaries that consist of
strongly magnetic white dwarf (WD) primaries and red dwarf (RD)
secondaries. Polars were first recognized in 1976 with the discovery
of circular polarization in AM Her (Tapia 1976, 1977).
Magnetic fields play a crucial role in determining
the system's parameters. The field strength of the primary is
so high that the material flowing from the companion does not form an
accretion disk around the WD, but is guided
along the field lines to an accretion column that forms near the magnetic pole of the primary.
The flux distribution from the column consists of
hard X-ray bremsstrahlung, an approximately
blackbody spectrum in the UV and soft X-ray and cyclotron emission, which is the primary source of optical radiation.

The brightest polar AM Herculis (RX J1816.2 +4952 $\equiv$ EUVE J1816+49.8 $\equiv$ 3U 1809+50 $\equiv$ H
1816+49) was classified as a cataclysmic variable (CV) by Berg \& Duthie (1977).
It is not an eclipsing binary despite the observed large-amplitude minima in the
light curves. The system has been studied for different wavelengths. Each study revealed different characteristics of the polar.
The system shows long-term, non-periodic variations where the brightness of the system varies by about 2 mag (see, Hessman, 2000),
known as high and low states. High and low states of polars are thought
to be due to the variation in the mass transfer rate from the RD to the WD.
In a recent study on the high and low states of the system by Wu \& Kiss (2008)
it was found that the magnetic field of the primary is a crucial parameter in
regulating these states.
Previously, Livio \& Pringle (1994) explained the observed low state of the system with the
starspots migrating under the inner Lagrangian point (L1).

\begin{figure*}
\includegraphics{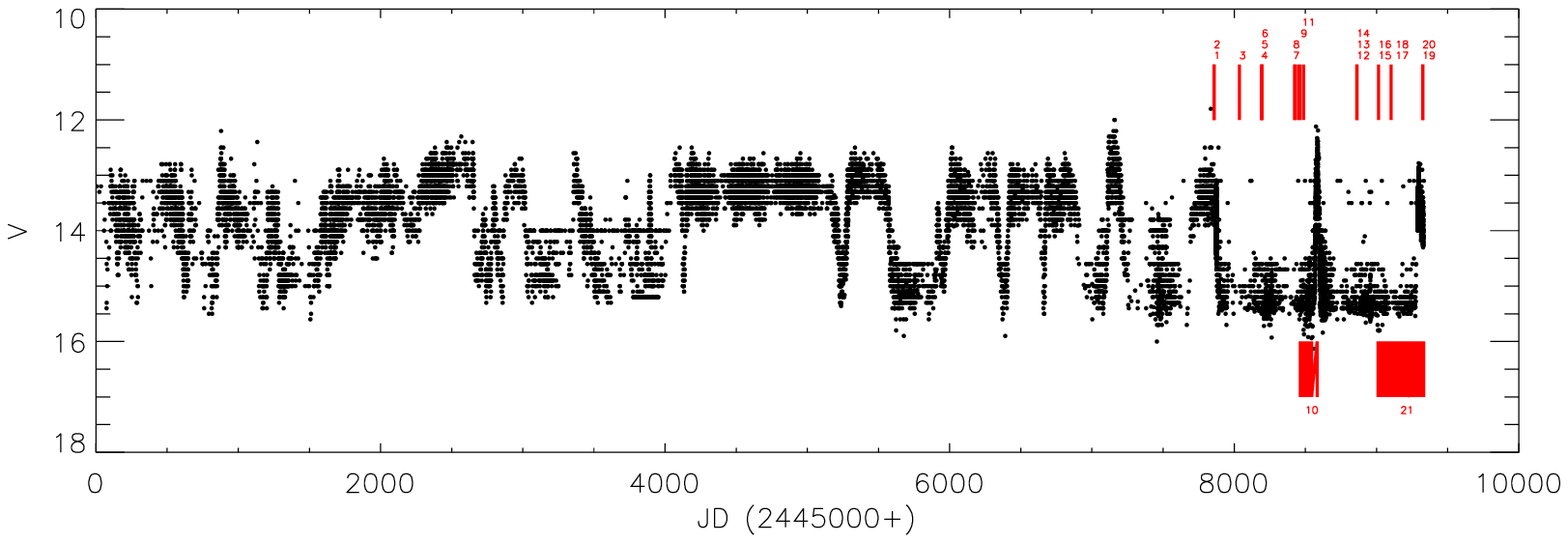}
\caption{Long (more than 25 years) optical light curve of AM Her from AAVSO data.
The times of RTT150 and ROTSE IIId are shown with vertical lines (see Table~1 for the data)}. (A color version of this figure is available in the online journal.)\label{f1}
\end{figure*}

\begin{figure*}
\includegraphics[angle=90]{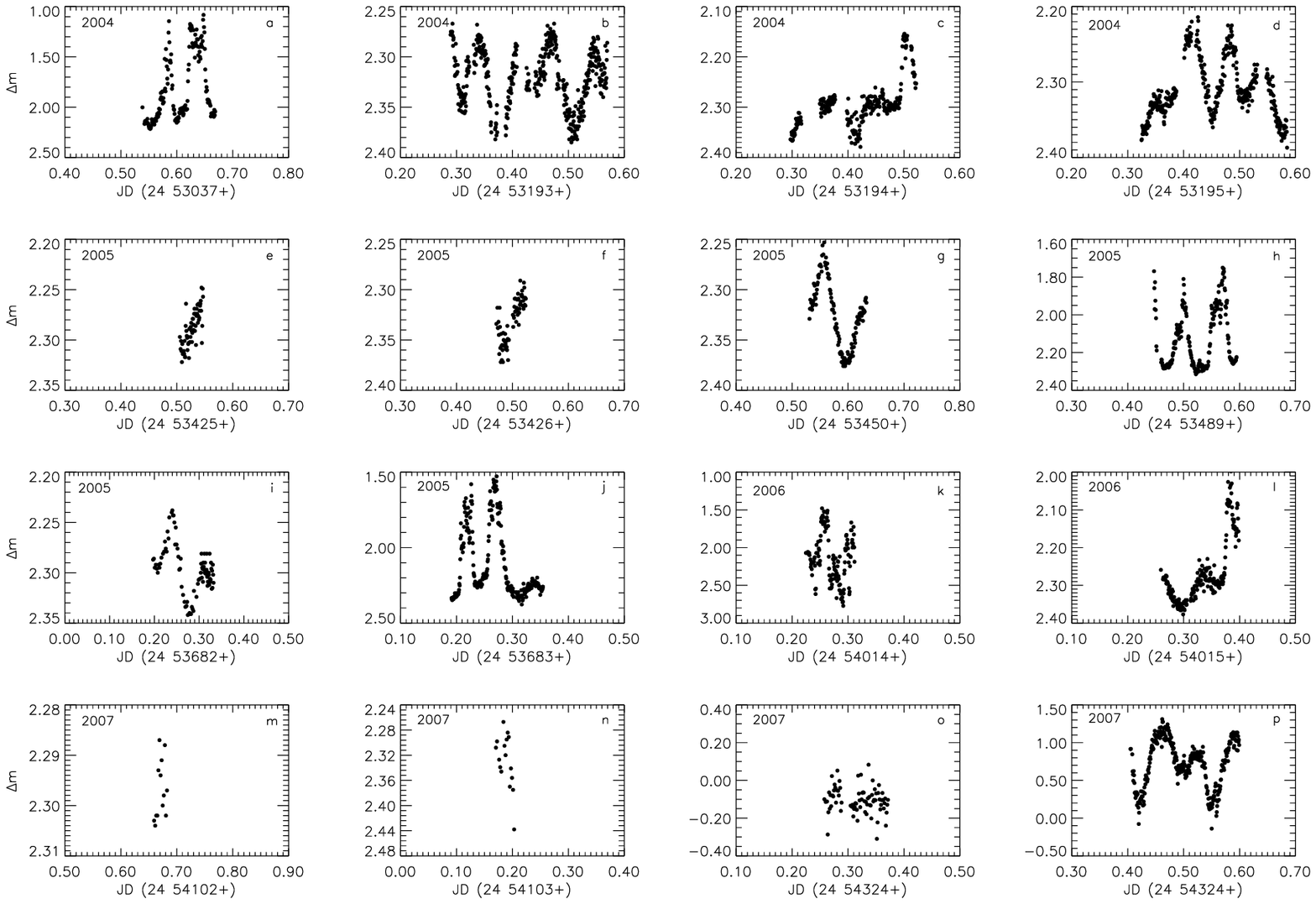}
\caption{Light curves of AM Her obtained between the period 2004--2007. All light curves are in R$_{c}$ except (O) that is obtained in I$_{c}$.}\label{f2}
\end{figure*}

Observational properties of polars generally depend on the
observed state and wavelength. For instance, short-term, low-amplitude variations in the X-ray/optical bands, known as flickering, are
detected when the system is in high
and intermediate states (King, 1989; Bonnet-Bidaud et al. 1991).
Different mechanisms have been suggested to explain these variations. Szkody \& Margon (1980), using the cross-correlation
functions of high state observations of AM Her,
reported the strong correlation in Johnson U, V, and $\lambda$4686 features
and discussed the ionizing radiation as a responsible mechanism for the observed
small-amplitude brightness variations. Another mechanism is the oscillation of the magnetic flux tubes (Tuohy et al. 1981). Larsson (1988), on the other hand,
proposed an oscillatory shock height model to explain
optical variations with periods, of a few seconds. King (1989) discussed that X-ray
irradiation of the accretion flow below the L1 point produces
oscillating ionization fronts. These ionization fronts modulate the
accretion rate through L1. The timescale of these oscillations is
the dynamical timescale near the L1, point which is about 8 minutes for AM Her.
Besides the high state, the low state of the system has also been the subject of
interest because of its physical properties and the poorly defined characteristics
that can change over time.

Complex and unpredictable observational properties of polars prompted us to
obtain long-term optical variation of AM Her.
Studying the light variations obtained over a long period of time is necessary to
see the complex structure.
The study of AM Her was carried out with the 1.5m RTT150 and ROTSE IIId (Robotic Optical Transient Search Experiment-IIId) telescopes of
the T\"UB\.ITAK National Observatory (TUG). The results are presented in Section 2. In Section 3 we analyze the observations and discuss them in Section 4.

\section{New observations and light variation}
Optical photometry of the system was obtained using the Russian--Turkish 1.5m telescope (RTT150)
over 19 nights between the period 2003-2007 (Table~\ref{amhertab1}).
All the images were obtained using the Andor CCD. The Andor CCD camera is
equipped with a set of Cousins (R$_{c}$, I$_{c}$) and Johnson ($V$) filters. During the data reduction a few comparison and check stars were chosen
in the same CCD frame including GSC~3533~1026 and GSC~3533~1021.
All CCD reductions were done with the IRAF\footnote{IRAF is distributed by
National Optical Astronomy Observatories, which is operated by the
Association of Universities for Research in Astronomy, Inc., under
cooperative agreement with the National Science Foundation, U.S.A. } package.
In Fig.~\ref{f1} the times of RTT150 and ROTSE IIId observations are indicated by vertical lines with the optical light curve of AM Her from the data of the American Association of Var,able Star Observers (AAVSO). Light variations of AM Her, obtained with RTT150, over long periods of time show that the light-variation amplitude changes over time.
The light curves, in Fig.~\ref{f2}, are plotted as a function of the Julian Date (JD) to make clear any possible variation between successive orbital phases (see, Kalomeni et al. 2005 for the light variation of the system obtained in 2003).

\begin{table}
\begin{center}
\scriptsize
\caption{Summary of the observations of AM Her with RTT150 and ROTSE IIId* telescopes. HJD$^{+}$ shows JD
start+2400000}\vspace{0.25cm} \label{amhertab1}
\begin{tabular}{lllcc}
\hline
Run     &    HJD$^{+}$ &  Filter & N$_{\rm{Obs}}$    & State \\
\hline
1       &    52858.46-52858.59         &R$_{c}$    &   162    &Intermediate   \\
2       &   52859.47-52859.59         &R$_{c}$    &  195      &Intermediate  \\
3       &    53037.55-53037.67         &R$_{c}$    &   148    &Active   \\
4       &    53193.29-53193.57      &R$_{c}$    &  444      &Low  \\
5       &   53194.30-53194.52     &R$_{c}$    &   261    &Active \\
6       &    53195.33-53195.59     &R$_{c}$    &   371    &Low \\
7       &   53425.51-53425.55         &R$_{c}$    &   61     &Low \\
8       &    53426.47-53426.53         &R$_{c}$    &   70     &Low \\
9       &    53450.53-53450.63      &R$_{c}$    &   129    &Low  \\
10*      &    53464.95-53582.84      &$-$        & 387  &Low + active + high  \\
11       &    53489.45-53489.60     &R$_c$    &   229    &Active \\
12      &    53682.20-53682.33        &R$_{c}$    &   90     &Low  \\
13      &    53682.20-53682.30        &V        &   51     &Low  \\
14      &   53683.19-53683.35        &R$_{c}$    &   266    &Active \\
15      &    54014.23-54014.31       &R$_{c}$    &   138    &Active  \\
16      &    54015.26-54015.40      &R$_{c}$    &   200    &Low  \\
17      &    54102.66-54102.68      &R$_{c}$    &   14     &Low  \\
18      &    54103.17-54103.20      &R$_{c}$    &   16     &Low  \\
19      &   54324.40-54324.56     &I$_{c}$    &   253    &High  \\
20      &  54324.40-54324.60      &R$_{c}$    &   297    &High  \\
21*     &    54012.22-54402.21  &$-$   &  1020   &Low + active + high \\
22*     &    54089.21-54573.53  &$-$   &  70      & High \\
\hline
\end{tabular}
\end{center}
\end{table}

\begin{figure*}
\includegraphics{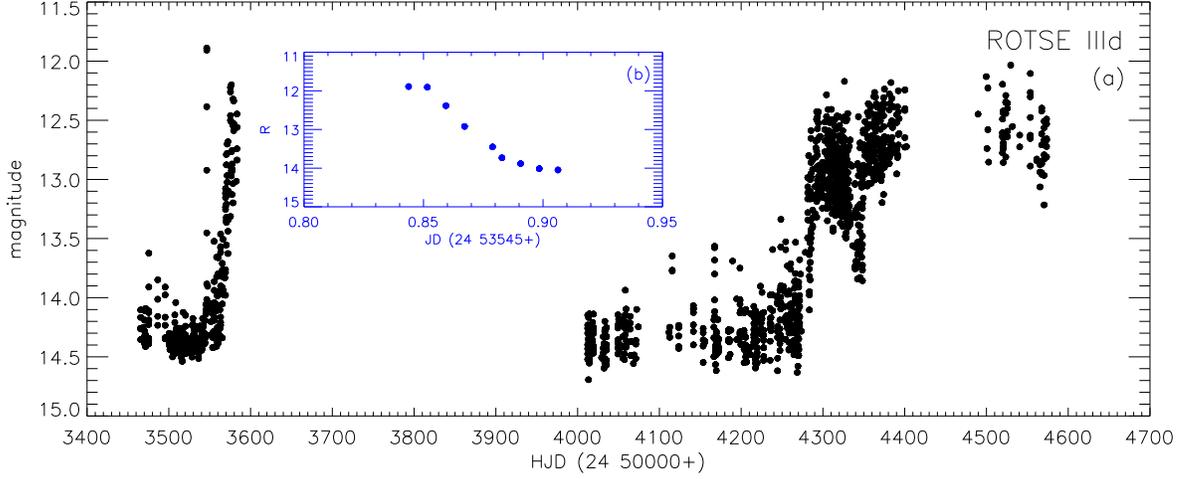}
\caption{(a) The figure shows long term light variation of AM Her spread over two hundred five nights obtained with the ROTSE IIId between the period 2005-2008, (b) the magnitude excess observed in HJD2453545}.\label{f3}
\end{figure*}

\begin{figure*}
\includegraphics{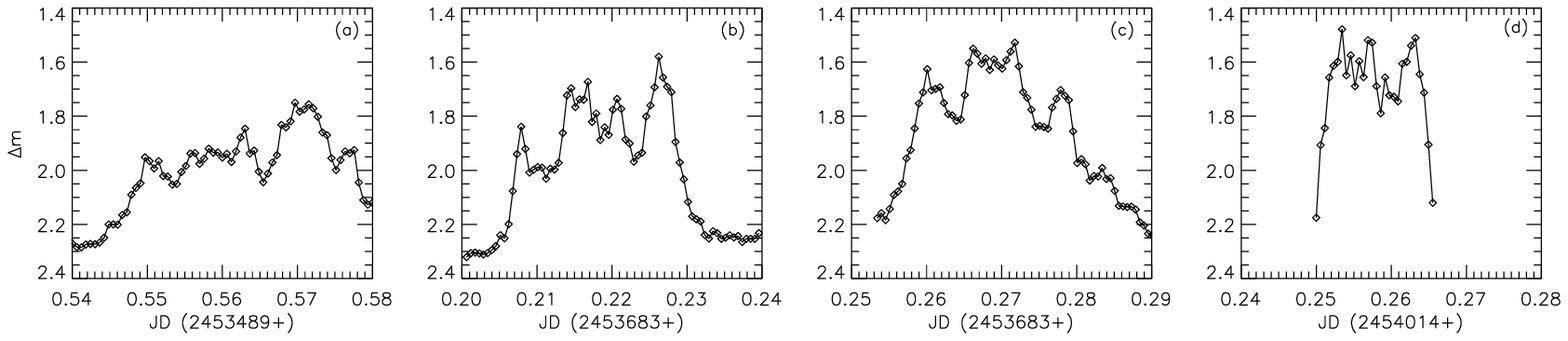}
\caption{Expanded
views of the R$_{c}$- band brightness variation during the low-state of
AM Her (a) April 28, 2005,
(b)--(c) November 8, 2005, and (d) October 5, 2006. }\vspace{+5cm}\label{f4}
\end{figure*}

ROTSE IIId\footnote{The ROTSE III system's details described in Akerlof et al. (2003).} was used to carry out long-term variation of the system  (Table~\ref{amhertab1}). Fig.~\ref{f3}a shows long-term light variation of AM Her spread over 205 nights, obtained with the ROTSE IIId, between the period 2005-2008. During the low state two noticeable brightness variations have been detected. Low-state observations of AM Her exhibit rather weak brightness variations with respect to the high and intermediate states.
During the low-state the mass transfer from the secondary is thought to decrease or cease.
If the accretion is almost negligible then the characteristic features of the component can exhibit themselves in
observed light curves (Kafka et al. 2005). Therefore, because the secondary is a late-type main-sequence star, we can expect to detect
stellar activity in the low state. Such a brightness variation in AM Her was observed by Shakhovskoy et al. (1993). They reported an approximately 2 mag flare event with a 20 minutes duration (see also, Bonnet-Bidaud et al. 2000). We detected similar events of 2.14 mag for 1 hour and 20 minutes in 2005 with ROTSE IIId (Fig.~\ref{f3}b). Following this variation ROTSE IIId detected a 1.37 mag excess in about 30 minutes on July 20, 2005 (see, Table~\ref{amhertab3}). During AM Her's low-state relatively small amplitude flaring with amplitudes of 0.2-0.6 mag and lasting 15-90 min was reported by Kafka et al. (2005). Likewise, during the observing runs performed between 2004 and 2006 similar variations were detected (Table~\ref{amhertab3}, Fig.~\ref{f4}).
Magnitude excesses, owing to the possible flare events, with respect to the quiescence level are shown in Table~\ref{amhertab3}. In the RTT150 observations we have also determined times of minima, derived by the Kwee--van Woerden method (Kwee \& van Woerden, 1956) and for the asymmetric minima by freehand curve, from the individual light curves. In Table~\ref{tab4} the times of minima are shown. The errors in times of minima are of the order of $0^\textrm{d}.0002-0^\textrm{d}.0007$.

\begin{table}
\begin{center}
\scriptsize \caption{Magnitude variations for AM Herculis. HJD$^*$ shows JD
start +2400000}\label{amhertab3}
\begin{tabular}{llllllll}
\hline
HJD$^*$      &Filter        & Phase   & magnitude excess
\\
\hline
53037.5712                    &R$_{c}$&          0.1&           0.7  \\
53037.6200                    &R$_{c}$&          0.5&           0.71      \\
53194.4846                    &R$_{c}$&          0.3&           0.12      \\
53489.5381                    &R$_{c}$&          0.8&           0.5\\
533545.8400                   &-&              0.52&          2.14\\
53571.8293                    &-&              0.01&          1.37\\
53683.2006                    &R$_{c}$&          0.8&           0.45\\
53683.2484                    &R$_{c}$&          0.2&           0.7\\
54014.2494                    &R$_{c}$&          0.56&          0.7\\
\hline
\end{tabular}\\
\end{center}
\end{table}

\section{Period Change Analysis}

Times of minima in AM Her, as well as in other polars, is known to offset (e.g. Bailey et al. 1993). The nature of the observed shift in times is poorly understood. However, if the orbital period of the system is determined accurately and the WD is synchronized with the orbital period, then these shifts generally attributed to the oscillation of the magnetic pole (Bailey \& Axon, 1981; Bailey et al. 1993). Any variation in the mass accretion rate alters the accretion geometry. In this case while the location of the pole remains fixed the position of the spot with respect to the pole changes (Cropper 1989).
On the other hand, the orbital periods of interacting binaries are known to change because of different processes. One of these is the mass transfer between the components. In AM Her systems, the RD component loses mass to the primary WD star. In semidetached binary systems, the displacement in minima times causes a parabolic variation, either upward or downward, in the difference between the observed (O) and calculated (C) time of minima diagram. Thus, analysis of the observed times of minima is important to determine any variation, due to mass transfer, in the orbital period. Unfortunately, there are almost no studies done on (O--C) variation of other polars in the literature. Previous (O--C) studies of AM Her were performed by Young \& Schneider (1979) and Mazeh et~al. (1986). The study of Young \& Schneider (1979) shows no evidence for any continuous period variation. The latter study by Mazeh et al. (1986) shows a downward curved parabola with $\dot P/{P}$=$-5\times10^{-14} $s$^{-1}$. We collated the additional minima times obtained since then with those obtained in this study (Table~\ref{tab4}) to revise the (O--C) variation. We assigned the same weight for all minima points during analysis. The starting epoch for the primary minimum was adopted from the Szkody \& Brownlee, (1977).
The (O--C) diagram of AM Her constructed with an initial light element
can be represented by the relation,
\begin{equation}
\begin{array}{l}
HJD\,\textrm{Min}I=24\,443014.7136(2)   +0.128927048(2)\times E \\
~~~~~~~~~~~~~~~~~~~~~~~~~~~~~~~~~~~~~~~~~+1.33(20)\times10^{-12}\times E^2\label{amher1}.
\end{array}
\end{equation}

\begin{figure}
\includegraphics{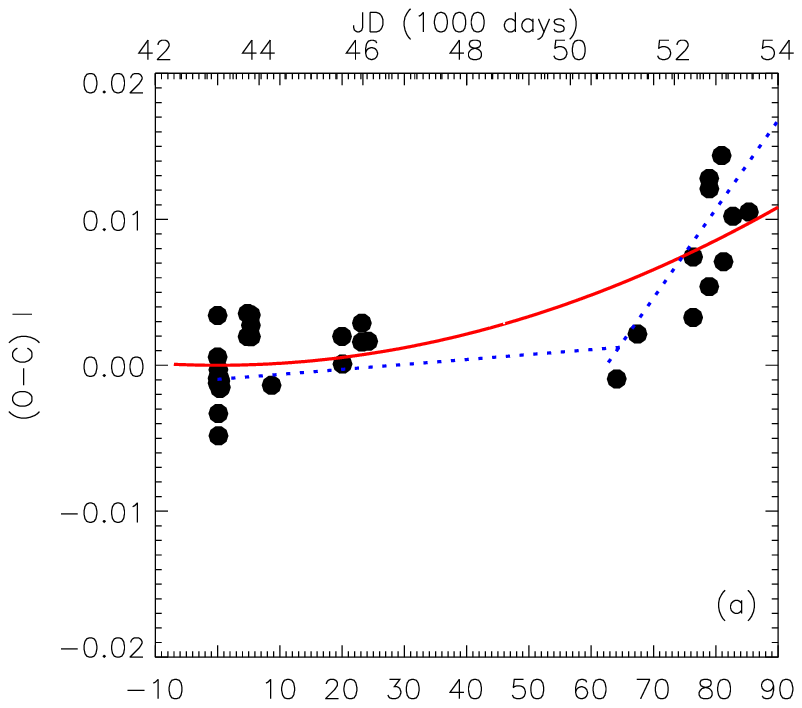}\\
\includegraphics{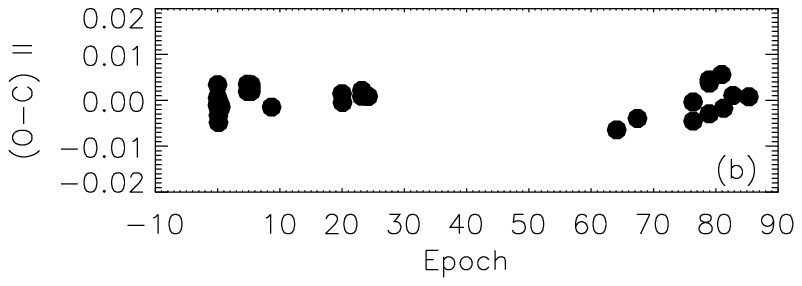}
\caption{a) Observed--calculated times of minima of AM Her vs. epoch, and b) the residuals of parabolic variation vs. epoch (see text for details). (A color version of this figure is available in the online journal.)}\label{f5}
\end{figure}

The (O--C) variation can be represented both by a parabola and with two broken lines.
First, if the long-term variation fits to a parabola, as is expected for binaries where mass transfer take place, the solid curve in Fig.~\ref{f5}a is for the best solution to the (O--C) variations. Then, the resulting solution with a quadratic term is an indication of the effect of mass transfer in AM Her. The mass transfer in polars is from the less massive secondary to the more massive primary WD star. This increases the orbital period in the conservative case and the resulting (O--C) curve is a parabola with a positive quadratic term. On the other hand, as a result of the nonconservative mass transfer and magnetic activity of the late-type component, mass loss from the system may also occur
in polars. However, the upward curving parabolic variation in the (O--C) diagram of AM Her indicates that the dominant effect is the mass transfer from the RD to the WD. If this variation fits to a parabola then the corresponding rate of period increase is $dP/dt=7.5(1.2)\times10^{-9}\,\ $days yr$^{-1}$ with a conservable mass transfer rate of $\dot M = {\dot P}/{3P} [{M_{1}M_{2}}/({M_1-M_2})]=7.6(2.3)\times10^{-9} \rm{M_{\odot}}$ yr$^{-1}$, for a WD with a mass 0.88$M_{\odot}$ (Bailey et al. 1988), which agrees with the maximum mass transfer rate given by Hessman et al. (2000) assuming that this variation is caused by stellar spots. On the other hand, the $P/ \dot P$ value is of the order of another recently studied polar (Andronov \& Baklanov, 2007). We can also fit the data with two broken lines. One of them is horizontal while the other one indicates an increase in the period. Nevertheless, observations in the next decades are necessary to clarify the true shape of the (O--C) diagram.

The accretion luminosity of the WD is $L=-GM_{1}\dot M/R_{1}$ where $M_{1}$ is the WD mass, $R_{1}$ is the radius of the WD, and $\dot M$ is the mass accretion rate.
Using the $\dot M$ estimated from the possible parabolic (O--C) variation, we can determine the accretion luminosity to derive the Alfv\'{e}n radius. The  Alfv\'{e}n radius is given by (Frank et al. 2002)
\begin{equation}
%\begin{array}{1}
r_{\mu} = 2.9\times10^8M_1^{1/7}R_{6}^{10/7}L_{37}^{-2/7} B_{12}^{4/7},
\label{amher3}
%\end{array}
\end{equation}
where $R_6$ is the radius of the WD in units of $10^6$cm, $L_{37}$ its luminosity in units of $10^{37}$ erg s$^{-1}$ and $B_{12}$ is the surface magnetic field strength in unit of $10^{12}$ G. The estimated Alfv\'{e}n radius for the adopted value of $M=0.88M_{\odot}$ is $2.02\times10^{10}$cm .

AM Her shows evidence for brightness variations on a timescale of minutes (approximately 4.5 minutes; e.g., Bonnet-Bidaud et al. 1991).
One of the explanations for the origin of the quasiperiodic X-ray variations observed in AM Her is discussed by Tuohy et al. (1981). They presented the oscillation of the magnetic flux tubes, through which matter flows to the stellar surface, as a possible mechanism for these variations. If these oscillations occur at the point where the matter channeled by the magnetic field, they can lead to the quasi-periodic variations in the accretion rate. The timescale for an Alfv\'{e}n wave to cross the magnetosphere is
\begin{equation}
%\begin{array}{l}
P_{osc}(r) = 2\times10^{-3} r_8^{11/4} L_{34}^{1/2}f_{-2}^{-1/2}M_1^{-3/4}R_{8}^{-2}B_7^{-1} \,\, \textrm{s},
\label{amher4}
%\end{array}
\end{equation}
where $r_8$ is the radius where the flowing matter is channeled and quasiperiodic variations may arise, $f$ is the fraction of stellar surface where the accreting matter flows. Adopting  $f=9.1\times10^{-3}$ as the characteristic dimensionless size of the flow (Hessman, 2000), we find the related $r$ on the timescales of interest as $2.03\times10^{10}$cm. The $r$ estimated for the 4 minutes oscillations agrees very well with the estimated Alfv\'{e}n radius. This indicates that this model adequately describes the observed brightness variations.

The Lagrangian radius of the WD is $R_{L1}/a=0.5-0.227\,$log$q$ (Plavec \& Kratochvil, 1964) with $q$ the mass ratio of the components and $a$ the orbital separation. For $q=0.31$ and $a=7.85\times10^{10}$cm, this yields $R_{L1}=4.8\times10^{10}$cm and $r_{\mu}\approx0.42R_{L1}$. Therefore, as is expected both the magnetospheric radius and the radius where the matter channeled toward the WD are smaller than the Lagrangian radius (Ferrario et al. 1989). The results obtained in this study are in agreement with the results presented by Bonnet-Bidaud et al. (1991). They discussed that their observed 270 s variations correspond to a radius $r_{\mu} \approx 2.1\times10^{10}$cm for AM Her. The flickering timescale depends on $\dot M$ and $f$ since $M_1$, $R$ and $B$ cannot change on the timescales of 3-8 minutes. Hence, as Bonnet-Bidaud et al. (1991) reported, any inhomogeneities in the accretion matter can produce these brightness variations on the timescales of interest.

\section{Results and Discussion}

AM Her-type systems show large-amplitude variations over years (Fig.~1-3)
as well as short-term low-amplitude variations.
In this study, five years of observations obtained in both states of the system are presented. Low-state photometry reveals weak orbital modulation but occasionally flaring-type variability of the secondary. Large flare events, as well as smaller amplitude flares, are detected.
Three of the nine magnitude excess events detected fell within the primary minimum and one within the secondary minimum. Durations of brightness variations range from tens of minutes to an hour.
Flickering with an amplitude of 0.01--0.60 mag occurs on a timescale of at least a few minutes.

We have obtained a total of nine times of minima, using them we could perform a period analysis of the system. The times of minima in AM Her show shift, which is, generally, assumed due to obscuration of the post-shock radiation of the main accretion column (Bailey et al. 1993). These changes are thought to be responsible for the observed shift in (O--C) diagram. On the other hand, AM Her systems are described with a mass-donating RD and a magnetic WD star, therefore, they are classified as semidetached binaries. Semi-detached binaries known to show a parabolic variation in the (O--C) diagram due to the mass transfer between the components. Using the available times of minima, we find for the first time, an evidence for an upward parabola general property observed in semi-detached binaries where the matter flows from less massive to the more massive. Using this variation we derive an orbital period evolution time of about 1.7$\times10^7$ yr. In addition, the (O--C) variation can be described with two broken lines.
A total of 30 years times of minima are collated. This time is quite long to see any variation in the (O--C) diagram. However, it is apparent from  Fig.~\ref{f5}a that the upward parabola is not so clear as seen in binaries with nondegenerate components (e.g., Kalomeni et al. 2007). If the period variation is due to the conservative mass transfer, the mass transfer rate between the components is $\dot M = 7.6\times 10^{-9}M_{\odot}yr^{-1}$. This mass transfer rate is small in comparison with the binaries with non-degenerate components (\emph{ibid}). Using timescales for the gravitational radiation and magnetic braking (Yakut et al., 2008)
we have calculated the gravitational radiation timescale and
magnetic braking timescale of AM Her about 7 Gyr and 1.3 Gyr, respectively.
The mass accretion rate timescale of the system is much less than these. Thus
the gravitational radiation is not as important as the other mechanisms for AM
Her. On the other hand, if the orbital period of AM Her were half of its present
value then gravitational radiation and magnetic braking would be much more
important. At periods of 1h, 2h, and 8h (e.g., V1309 Ori) the gravitational radiation
timescales are about 0.4 Gyr, 2.4 Gyr, and 96 Gyr, respectively.
In AM Hers, the binary geometry, period change ratio, angular momentum loss mechanisms, etc. can all
change because of the strong magnetic field of the WD and the interaction between components' magnetic field
(see also Wickramasinghe \& Wu, 1994; Webbink \& Wickramasinghe, 2002).

\begin{table}
\begin{center}
\scriptsize \caption{The times of minima of
 AM Her in HJD* (HJD - 2\,400\,000).}\label{tab4}
\begin{tabular}{llllll}
\hline
HJD$*$   & Passband          & Ref.   &HJD$^*$     & Passband            & Ref.
\\
\hline
43014.71266	&	V		& 1&	44133.02540	   &	V		& 4	 \\
43014.84127	&	V		& 1&	45591.32259 	&	V		& 5 \\
43015.74554	&	V		& 1&  45600.34559	&	V		& 5   \\
43015.87731	&	V		& 1 &		46000.2788	&	V		& 5   \\
43031.72862	&	V 		& 1&	46001.31151	&	V		& 5 \\
43031.86055	&	V		& 1	&   46132.55800	&	V		& 5\\
43032.6336  	&	I		& 2&  51277.5182	&	-		& 6 	\\
43033.661  	&	I		& 2	&    51708.45993      &    980-1180{\AA} &7 \\                            	
43062.5439  	&	I		& 2&   52858.5548 	&	Rc		&8 \\                               	
43062.8024    &	V		& 1&     	52859.5259	&	Rc		&8\\
43069.6354      &	I		& 2	&   53193.3160	&	Rc		&8  \\
43083.5591      &	I		& 2	&   53193.5168	&	Rc		&8 \\
43635.88762 	&6900-7400{\AA}	& 3	&   53195.4500	&	Rc		&8  \\                   	
43636.78853	&	8200-8700{\AA}		& 3	&53450.5989	&	Rc		&8  \\
43704.73386	&	8000-8350{\AA}		& 3 & 53489.5276	&	Rc		&8 	\\
43704.86349	&	8000-8350{\AA}		& 3 &   53683.2436	&	Rc		&8\\
43705.89342	&	7500-7800{\AA}		& 3 &54015.2955	&	Rc		&8	 \\
\hline
\end{tabular}
\end{center}
{References for Table ~\ref{tab4}: 1-Szkody \& Brownlee (1977), 2-Olson (1977), 3-Young \& Schneider (1979),
4-Young et al. (1981) based on Bailey \& Axon (1981) observations, 5-Mazeh et al. (1986),
6-Safar \& Zejda (2002), 7-Hutchings et al. 2002 (the average minimum time is used), 8-This study.}
\end{table}

\acknowledgements

We are indebted to the E. R. Pek\"unl\"u, C. A. Tout and anonymous referee for their valuable comments.
We would thank to J. Eldridge for the reading final version
of the MS and V. Keskin and \"U. K{\i}z{\i}lo\v{g}lu for their support in ROTSE data process and reduction. We acknowledge that in this study we have used the data from the AAVSO Database.
This work has been partly supported by T\"UB\.ITAK National Observatory and T\"UB\.ITAK-B\.IDEB.

\end{document}